\documentclass[journal]{IEEEtran}

\usepackage{cite} 
\usepackage{epsfig,amssymb}   
\usepackage{graphicx}  
\usepackage[cmex10]{amsmath}
\usepackage{verbatim}
\usepackage{bbm}
\usepackage{psfrag}
\usepackage[usenames]{color}
\usepackage{multirow}
\usepackage{multicol}
\usepackage{stfloats}

\newcommand{\p}[1]{\mathop{\mbox{\it p} } }

\renewcommand{\vec}[1]{\ensuremath{\boldsymbol{#1}}}
\newcommand{\be}{\begin{equation}}
\newcommand{\ee}{\end{equation}}
\newcommand{\ba}{\begin{array}}
\newcommand{\ea}{\end{array}}
\newcommand{\bea}{\begin{eqnarray}}
\newcommand{\eea}{\end{eqnarray}}
\newcommand{\bean}{\begin{eqnarray*}}
\newcommand{\eean}{\end{eqnarray*}}

\newcommand{\rmr}{^{\rm r}}

\definecolor{white}{rgb}{1,1,1}

\newtheorem{theorem}{Theorem}

\begin{document}

\title{Optimal Transmit Filters for ISI Channels under Channel Shortening Detection}

\author{Andrea Modenini,~\IEEEmembership{Student~Member,~IEEE,} Fredrik~Rusek, and~Giulio~Colavolpe,~\IEEEmembership{Senior~Member,~IEEE}\\
\thanks{Submitted: May 23, 2013. Revised: \today.
A. Modenini and G. Colavolpe are with Universit\`a di Parma, Dipartimento di Ingegneria dell'Informazione, Parco Area delle Scienze, 181A, 43124 Parma, 
Italy, e-mail: giulio@unipr.it, modenini@tlc.unipr.it. F. Rusek is with Lund University,  box 118, 22100  Lund, Sweden, e-mail: fredrik.rusek@eit.lth.se. 
}
\thanks{The work of F.~Rusek was supported by SSF through the distributed antenna project.
}
\thanks{The paper was presented in part at the IEEE Intern. Conf. Commun. (ICC'13), Budapest, Hungary, June 2013.}
}

\markboth{IEEE Transactions on Communications}{Submitted paper}

\maketitle 

\begin{abstract}
We consider channels affected by intersymbol interference with reduced-complexity, mutual information optimized, channel-shortening detection.  For such settings, we optimize the transmit filter, taking into consideration the reduced receiver complexity constraint. 
As figure of merit, we consider the achievable information rate of the entire system and  with functional analysis, we establish a general form of the optimal transmit filter, 
which can then be optimized by standard numerical methods. 
As a corollary to our main result, we obtain some insight of the behavior of the standard waterfilling algorithm for intersymbol interference channels. 
With only some minor changes, the general form we derive can be applied to  multiple-input multiple-output channels with intersymbol interference.
To illuminate the practical use of our results, we provide applications of our theoretical results by deriving the optimal shaping pulse of a linear modulation transmitted over a bandlimited additive white Gaussian noise channel which has  possible applications in the faster-than-Nyquist/time packing technique.
\end{abstract}


\section{Introduction}
The intersymbol interference (ISI) channel has played a
central role in communication theory for several decades. It
has been heavily researched, and today most of its fundamental
 properties are known. The capacity of the ISI channel was for example
derived by Hirt back in 1988 in \cite{Hi88}, and it was shown that Gaussian inputs
in combination with the classical waterfilling 
algorithm achieves capacity. In practice, Gaussian channel inputs are
not very common and discrete inputs are typically
preferred. In this case the ultimate communication limit was found
in the early 2000s through a series of papers
\cite{ArLo01,PfSoSi01,ShSi01,PfSoSi07,ArLoVoKaZe06}. Further results on capacity
properties of ISI channels include Kavcic's elegant method
\cite{Ka01} to achieve the capacity of the ISI channel with discrete inputs through a
generalized version of the Arimoto-Blahut algorithm, and also Soriaga
et al.'s evaluation of the low-rate Shannon limit of ISI channels
\cite{SoPfSi03}. 

However, all of the above mentioned papers study ISI channels under
the assumption that the receiver can perform optimal maximum-likelihood (ML) or maximum-a-posteriori (MAP) detection. Let $L_{\mathrm{H}}+1$ denote the number of taps in the channel impulse response. Forney showed in 1972 \cite{Fo72b} that optimal ML/MAP-detection can be performed by searching a trellis whose number of states is $U^{L_{\mathrm{H}}}$, where $U$ is the cardinality of the employed constellation. The number of trellis states will be considered in the following has a measure of the receiver complexity. In many practical
scenarios $L_{\mathrm{H}}$ is far too long for
practical implementation of optimal ML/MAP detection. This observation spurred significant
research efforts to reduce the computational complexity of the MAP/ML algorithm (e.g., see~\cite{FeBaCo07,CoFePi11} and references therein)
or to investigate when a properly designed linear equalizer has the same diversity order of the optimal detector (e.g. see~\cite{MaZh08,SoLe11} and references therein). An alternative promising approach was channel
shortening pioneered by Falconer and Magee in 1973 \cite{FaMa73} and further investigated by several researchers (e.g., see~\cite{Fr76,Be78,SuEdOdErKoOr94,AlCi96,Al01,AlAlAl00,VeEr07,BaAl02,DaGeSl11,DaGeSl11b}). Traditionally,
channel shortening detectors were optimized from a
minimum mean-square-error (MMSE) perspective. However, minimizing the mean-square-error does not directly correspond to achieving the highest information rate (in the Shannon sense) that can be supported by a shortening
detector. Recently, the achievable rate of channel-shortening detectors was optimized in
\cite{RuPr12} by utilizing the framework of  mismatched mutual
information \cite{MeKaLaSh94,GaLaTe00}.  The result of \cite{RuPr12} is a closed-form expression of the achievable information rate (AIR) of an ISI channel
with Gaussian inputs and an optimized channel-shortening detector that
considers the channel memory to be $L<L_{\mathrm{H}}$ taps long, where $L$ is a
user-defined parameter. 

In this paper, we extend \cite{RuPr12} by designing a proper transmit filter 
to be employed jointly with a channel-shortening detector\footnote{As in \cite{RuPr12}, with the term ``channel-shortening detector'' we mean a detector based on a proper linear filter (the channel shortener) plus a suboptimal reduced-complexity trellis-based detector with proper branch metrics designed for a target channel response of length $L<L_{\mathrm{H}}$. With ``optimal'' channel-shortening detector we mean that proposed in \cite{RuPr12} which is optimal from the point of view of the maximization of the achievable information rate.
} with the aim of further improving
the achievable information rate. In other words,
\textit{we consider to adopt, at the receiver side, a channel-shortening detector and then solve for the optimal transmit filter to be used jointly with it.}
When the use of the optimal full-complexity receiver is allowed, the answer to this question is the classical waterfilling processing. We are generalizing the waterfilling concept to the case of reduced-complexity channel-shortening detectors, i.e., we
essentially redo Hirt's derivations, but this time with the practical
constraint of a given receiver complexity.

Our results are not as conclusive as in the
unconstrained receiver complexity case. With functional analysis, we
can prove that, for real channels, the optimal transmit filter has a frequency response
described by $L+1$ real-scalar values. In general, for complex channels,
the optimal transmit filter is described by $L+1$ complex scalar values.
The transmit filter optimization thereby becomes a problem of finite dimensionality, and a
numerical optimization provides the optimal spectrum. Note that, in
practice, $L$ is limited to rather small values and $L=1$ is an
appealing choice from a complexity perspective. This essentially leads to very effective numerical optimizations.


The rest of the paper is organized as follows. In Section \ref{sysmod}, we lay down the system model and formulates the problem that we intend to solve. 
In Section \ref{solution}, we derive a general form of the frequency response of the optimal transmit filter. 
In Section \ref{sec:continuous}, we derive, by using the same framework, 
the optimal transmit filter for 
multiple-input multiple-output channel (MIMO) affected by ISI (MIMO-ISI), and the optimal shaping pulse
for a transmission over a bandlimited additive white Gaussian noise (AWGN) channel.
Numerical examples and properties of the numerical optimization are given in Section \ref{numres}. 
Finally, Section \ref{conclusion} concludes the paper.

\section{Preliminaries} \label{sysmod}
In this section we give the system model, lay down the fundamentals of
channel shortening receivers and their optimization, and formulates
the problem that will be solved.
\subsection{System Model}
Let us consider the transmission of the sequence of symbols $\vec{a}=\{a_k\}$ over a discrete-time channel with model\footnote{For simplicity of exposition, we refer here to this discrete-time model of a channel with finite ISI. We will discuss later the case of a continuous-time, bandlimited AWGN channel.}
\be \label{chmod}
y_k=\sum_{\ell=0}^{L_{\mathrm{H}}} a_{k-\ell} h_{\ell} + w_k,
\ee
where $\vec{h}=\{h_k\}_{k=0}^{L_{\mathrm{H}}}$ is the channel impulse response, assumed time-invariant and of finite length, and
$\vec{w}=\{w_k\}$ are independent and identically distributed complex Gaussian
random variables, with mean zero and variance~$N_0$---note that
bold letters are used for vectors. 
This system is studied under the assumption of ideal channel
state information (CSI) at both transmit and receive side, that is, perfect knowledge of the coefficients and the
noise variance.
The symbol vector $\vec{a}$ is a precoded version of the information
symbols $\vec{u}=\{u_k\}$,

\begin{equation}
	\vec{a}=\vec{u}\star \vec{p}\,,
\end{equation}
where ``$\star$'' denotes convolution and $\vec{p}$ is a transmit
filter subject to the power constraint $\sum_{k} |p_k|^2=1$ and with continuous spectrum $|P(\omega)|^2$, where $P(\omega)$ is  the discrete time Fourier transform (DTFT) of the vector $\vec{p}$.
Taken together, the received signal can be expressed as
\be \label{DeepEuroCrisis} \vec{y}=\vec{v}\star \vec{u}+\vec{w},\ee
where $\vec{v}=\vec{h}\star \vec{p}$. It is convenient to assembly the
presentation on matrix notation, so that (\ref{DeepEuroCrisis})
becomes
\be \nonumber \vec{y} = \vec{V} \vec{u}+\vec{w},\ee
where $\vec{V}$ is a convolutional matrix formed from the vector
$\vec{v}$, and $\vec{y}$, $\vec{u}$ and $\vec{w}$ are now column
vectors of appropriate sizes. 
Assume that the combined channel-precoder response $\vec{v}$ has
$K+1$ non-zero taps. The complexity of MAP sequence (implemented through the
Viterbi algorithm) and symbol detection (implemented through the BCJR algorithm) is
$\mathcal{O}(U^K)$ per  symbol, where $U$ is the cardinality of the employed
alphabet. Falconer and Magee's idea was to reduce this
complexity by a linear filtering 
\be \nonumber \vec{r}=\vec{y}\star
\vec{q}=(\vec{v}\star\vec{q})\star \vec{u}+(\vec{w}\star\vec{q}).\ee
Then, a Viterbi/BCJR algorithm follows assuming a \textit{target
  response} $\vec{t}$ of $L+1$ taps ($L\leq K$), and working on a trellis with $U^L$ states. Presumably, the target response
$\vec{t}$ roughly equals the $L+1$ strongest taps of
$(\vec{v}\star\vec{q})$, but there must not be an exact match if it
turns out that it is not optimal to do so. In matrix notation, this
procedure can be viewed as if the receiver decodes on the basis of a
mismatched conditional probability distribution (pdf)\footnote{By
$\vec{T}$ and $\vec{Q}$ we mean the convolutional matrices
formed from the vectors $\vec{t}$ and $\vec{q}$, respectively.}
\be \label{TotalGarbageCrisisInTheSouth}
\tilde{p}(\vec{y}|\vec{u}) \propto
\exp\left(-\frac{\|\vec{Q}\vec{y}-\vec{T}\vec{u}\|^2}{N_0}\right)
\ee
instead of the actual conditional pdf
\be \nonumber
p(\vec{y}|\vec{u}) \propto
\exp\left(-\frac{\|\vec{y}-\vec{V}\vec{u}\|^2}{N_0}\right).
\ee

Two questions now emerge: (1) For a given target response $\vec{t}$,
how should the linear filter $\vec{q}$ be selected? And (2) how should
the target response $\vec{t}$ be selected? These two questions kept
researchers busy for several decades, see
\cite{Fr76,Be78,SuEdOdErKoOr94,AlCi96,Al01,AlAlAl00,VeEr07,BaAl02,DaGeSl11,FaMa73}.
However, in all of those papers, the optimizations of $\vec{t}$ and
$\vec{q}$ was done with an MMSE cost function, which does not directly
correspond to the achievable information rate of the overall
system.\footnote{With ``overall system'', we mean the chain:
  prefilter-channel-reduced complexity receiver.}

The optimization for achievable information rate was completely solved
in \cite{RuPr12} under the assumption of Gaussian input symbols and by using a slightly more general model
for channel shortening. This generalization is now described.
By expansion of the exponent in (\ref{TotalGarbageCrisisInTheSouth})
we get
\bea \label{Ndrangheta}
\tilde{p}(\vec{y}|\vec{u}) &\propto&
\exp\left(-\frac{\|\vec{Q}\vec{y}-\vec{T}\vec{u}\|^2}{N_0}\right)\nonumber\\
&\propto& \exp\left(\frac{2\mathcal{R}\{\vec{u}^\dagger \vec{T}^\dagger\vec{Q}\vec{y}\}-\vec{u}^\dagger
    \vec{T}^\dagger \vec{T}\vec{u}}{N_0}\right),
\eea
where all terms independent of $\vec{u}$ have been left out. A MAP sequence
detector based on (\ref{Ndrangheta}) was proposed by Ungerboeck in
1974 \cite{Un74} and an algorithm for MAP symbol detection in 2005 by Colavolpe and Barbieri~\cite{CoBa05b}. In \cite{RuPr12}, a
reduced complexity channel shortening detector is obtained by
substituting in (\ref{Ndrangheta}) $\vec{T}^\dagger\vec{Q}$ with $(\vec{H}\rmr)^\dagger$ and
$\vec{T}^\dagger \vec{T}$ with $\vec{G}\rmr$. In
addition, the noise density $N_0$ is also absorbed into $\vec{H}\rmr$
and $\vec{G}\rmr$. This results in a mismatched conditional pdf of the
form
\be  \nonumber
\tilde{p}(\vec{y}|\vec{u}) =\exp\left(2\mathcal{R}\{\vec{u}^\dagger
  (\vec{H}\rmr)^\dagger \vec{y}\}-\vec{u}^\dagger
    \vec{G}\rmr \vec{u}\right).
\ee

While the front-end  $\vec{H}\rmr$ is
unconstrained, the matrix $\vec{G}\rmr$ must satisfy 
\begin{equation}
	G\rmr_{\ell k} = 0, \quad |\ell-k|>L \label{eq:const_G}
\end{equation}
in order to satisfy the reduced-complexity constraint.
The matrix $\vec{T}^\dagger \vec{T}$ in (\ref{Ndrangheta})
must be positive semi-definite, while no such constraint applies to
the matrix $\vec{G}\rmr$. Hence, a more general model than
(\ref{TotalGarbageCrisisInTheSouth}) for
channel shortening is obtained.
The AIR of a general mismatched receiver is
derived in \cite{GaLaTe00,MeKaLaSh94} and equals
\be \nonumber I_{\mathrm{AIR}} =  \lim_{N\to\infty} \frac{1}{N} \left[-\mathbb{E}_{\vec{y}}
\left[\log_2\left(\tilde{p}(\vec{y})\right)\right] +\mathbb{E}_{\vec{y},\vec{u}}
\left[\log_2\left(\tilde{p}(\vec{y}|\vec{u})\right)\right]\right],\ee
where $N$ is the number of input symbols (i.e., the length of the vector $\vec{u}$), $\mathbb{E}_{\vec{y}}$ denotes the expectation operator with
respect to the random variable $\vec{y}$ and
\be  \nonumber
\tilde{p}(\vec{y}) \triangleq \sum_{\vec{u}}\tilde{p}(\vec{y}|\vec{u})p_{\vec{u}}(\vec{u}).\ee
The rate  $I_{\mathrm{AIR}}$ is directly impacted by the choices
of $\vec{G}\rmr$ and $\vec{H}\rmr$. 
The optimization problem reads

\begin{equation}
	I_{\mathrm{OPT}}=\max_{\vec{G}\rmr, \vec{H}\rmr} I_{\mathrm{AIR}}, \label{eq:opt_problem_CS}
\end{equation}
under the constraints specified in (\ref{eq:const_G}).
Problem (\ref{eq:opt_problem_CS}) for a discrete alphabet is a hard task.
On the other hand, it can be solved in closed form  under the assumption that transmitted symbols are
independent Gaussian random variables~\cite{RuPr12}.
In this case of Gaussian inputs, closed-form expressions for $\vec{G}\rmr$, $\vec{H}\rmr$
can be found with the following algorithm:
\begin{itemize}
 \item Compute the sequence $\{b_k\}_{k=-L}^L$ as
	\bea
	b_k&=&\frac{1}{2\pi}\int_{-\pi}^{\pi}\frac{N_0}{|V(\omega)|^2+N_0} e^{jk\omega} \mathrm{d}\omega
	\nonumber \\
	&=&\frac{1}{2\pi}\int_{-\pi}^{\pi}\frac{N_0}{|H(\omega)|^2|P(\omega)|^2+N_0}e^{jk\omega}\mathrm{d}\omega. \nonumber\eea
	where $H(\omega)$ and $V(\omega)$ are the DTFT of $\boldsymbol{h}$ and $\boldsymbol{v}$.
 \item Compute the real-valued scalar
	\be \label{cc} c = b_0 -{\bf b}{\bf B}^{-1}{\bf b}^{\mathrm{\dagger}},\ee
	where ${\bf b}=[b_1,b_2,\ldots, b_{L}],$
	and ${\bf B}$ is $L\times L$ Toeplitz with entries $B_{ij}=b_{j-i}$.
  
 \item Define the vector $\boldsymbol{u}=  \frac{1}{\sqrt{c}} [1,\, -{\bf b}{\bf B}^{-1}]$ and find the optimal
       target response as
   \begin{equation}\nonumber
      G^r(\omega)= |U(\omega)|^2 - 1 \, .
   \end{equation}
 \item Finally, the optimal channel shortener is found as
    \begin{equation}\nonumber
      H^r(\omega)=\frac{ V(\omega)}{|V(\omega)|^2 + N_0 }(G^r(\omega)+1) ~~.
    \end{equation}
\end{itemize}
By using the optimal channel shortener and the target response $I_{\mathrm{OPT}}$ results to be
\begin{equation} \nonumber
I_{\mathrm{OPT}}=-\log_2(c) \, . 
\end{equation}


\subsection{Problem Formulation}
The problem we aim at solving is to maximize $I_{\mathrm{OPT}}$ over
the transmit filter $P(\omega)$, i.e., the DTFT of
$\vec{p}$. Thus, we have the following optimization problem at hand
\bea \label{ClownPrimeMinisters} &\min_{P(\omega)} c[P(\omega)]& \nonumber \\
&\mathrm{such\; that}&  \\
& \int_{-\pi}^{\pi} |P(\omega)|^2\mathrm{d}\omega =2\pi&\nonumber\eea
In (\ref{ClownPrimeMinisters}), we have made explicit the
dependency of $c$ on $P(\omega)$, but not on $N_0$ and $H(\omega)$,
since these are not subject to optimization. Since the starting point is the expression of the AIR when the optimal channel-shortening detector is employed, we are thus jointly optimizing the channel shortening filter, the target response, and the transmit filter, although for Gaussian inputs only. However, as shown in the numerical results, when a low-cardinality discrete alphabet is employed, a significant performance improvement is still observed (see also~\cite{RuPr12}).

\section{General Form of the Optimal Transmit Filter }
\label{solution}
The optimization problem (\ref{ClownPrimeMinisters}) is an instance of
calculus of  variations. We have not been able to solve it in
closed form, but we can reduce the optimization problem into an $L+1$
dimensional problem, which can then  efficiently be solved by standard
numerical methods. The main result of the paper is the following theorem.
\begin{theorem}
The optimal transmit filter with continuous spectrum for the channel $H(\omega)$ with a memory
$L$ channel-shortening detector satisfies

\begin{equation}
 |P(\omega)|^2 =\max\left(0,\frac{N_0}{\sqrt{|H(\omega)|^2}}\sqrt{\sum_{\ell=-L}^L A_{\ell}
      e^{j\ell \omega}}-\frac{N_0}{|H(\omega)|^2}\right)\,,\label{eq:Pw_opt}
\end{equation}
where $\{A_{\ell}\}$ are complex-valued scalar constants with Hermitian symmetry, i.e.,  
$A_{\ell}=A^*_{-\ell}$.
\end{theorem}

For a proof see the Appendix A.

\section{Interlude: Full Complexity Detectors}
Theorem 1 gives a general form of the optimal transmit filter to be used
for a memory $L$ channel shortening detector. By definition, it 
becomes the classical waterfilling filter when $L=K$.
Hence, it also provides
an insight to the behavior of the transmit filter for the classical
waterfilling algorithm. We remind the reader that $L_{\mathrm{H}}+1$ denotes the duration of the channel impulse
response and $K+1$ denotes the duration of the combined transmit
filter and channel response. We summarize our finding in the following
\begin{theorem}\label{th:memory}
Let $P(\omega)$ be the transmit filter found through the
waterfilling algorithm. Then, 
$$ K\geq L_{\mathrm{H}}.$$
\end{theorem}

For a proof, see the Appendix B.

Whereas the statement is trivial when the transmit filter and the channel
have a finite impulse response (FIR), the theorem proves that this fact holds
also when they have infinite impulse responses (IIR). 
Thus, for a FIR channel response, the waterfilling solution cannot contain any pole that cancels a zero of the channel, while, for IIR channels, 
the waterfilling solution cannot contain any zero that cancels a pole. Thus,
the overall channel cannot be with memory shorter than the original one.

Theorem 2 reveals the interesting fact that the waterfilling
algorithm trades  a rate gain for detection complexity. By using the
optimal transmit filter, a capacity gain is achieved, but the
associated decoding complexity (of a full complexity detector) must
inherently increase. Thus, with waterfilling, it is not possible to
achieve both a rate gain and a decoding complexity reduction at the
same time.

\section{Other practical applications of the optimal transmit filter}
Although we restricted our attention on the discrete-time ISI channel~(\ref{chmod}),
the same framework can be used to derive the optimal
precoder for other channels.
\subsection{MIMO-ISI Channels with perfect CSI}
Consider the MIMO-ISI channel
\begin{equation}
	\boldsymbol{y}_k= \sum_{\ell=0}^{L_H} \boldsymbol{H}_{\ell} \boldsymbol{a}_{k-\ell} + \boldsymbol{w}_k \,.\nonumber
\end{equation}
Without loss of generality, we assume that the channel is $N\times N$, i.e.,  matrices $\{\boldsymbol{H}_\ell\}_{\ell=0}^{L_H}$ have dimension $N\times N$
and $\{\boldsymbol{y}_k\}$,$\{\boldsymbol{a}_{k}\}$,$\{\boldsymbol{w}_k\}$ are column vectors $N\times1$.
In case $N\times M$ channels, they can be converted in a equivalent $N\times N$ channel by means of the QR decomposition\cite{RuPr12}. Channel shortening receivers for MIMO-ISI channels have been studied before, e.g., in \cite{Al01}, but here we optimize the receiver with respect to mutual information rather than an MMSE cost function as in \cite{Al01}.

\begin{figure}
	\begin{center}
		\includegraphics[width=1.0\columnwidth]{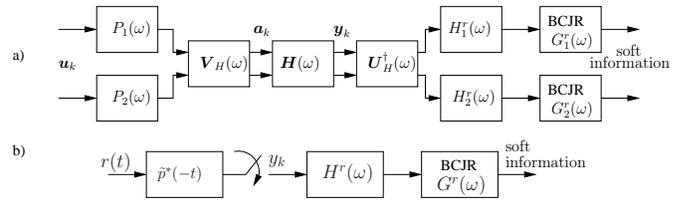}
		\caption{Block diagrams of a) the transceiver for $2\times 2$ MIMO-ISI channels and of b) the channel shortening detector for continuous-time AWGN channels.}\label{fig:bd_awgn}
	\end{center}
	\vspace{-6mm}
\end{figure}

The DTFT of $\{\boldsymbol{H}_\ell\}$, defined as $\boldsymbol{H}(\omega) = \sum_{\ell=0}^{L_H} \boldsymbol{H}_\ell e^{-j\ell\omega}$,
can be factorized by means of singular value decomposition (SVD) as
$$\boldsymbol{H}(\omega) = \boldsymbol{U}_H(\omega) \boldsymbol{\Sigma}(\omega) \boldsymbol{V}^{\dagger}_H(\omega)\,,$$
where $\boldsymbol{U}_H(\omega)$ and $\boldsymbol{V}_H(\omega)$ are unitary matrices and $\boldsymbol{\Sigma}(\omega)$ is a diagonal matrix
with elements $\Sigma_{n} (\omega)$.
By adopting the MIMO filter $\boldsymbol{V}_H(\omega)$ at the transmitter and the filter
$\boldsymbol{U}^{\dagger}_H(\omega)$ at the receiver, without any information loss we obtain $N$ independent parallel channels with channel
responses $\{\Sigma_{n} (\omega)\}_{n=1}^N$.
The transceiver block diagram is as shown in Fig.~\ref{fig:bd_awgn}a for the case $N=2$.
The objective function to be maximized is
\begin{equation} 
	I_{\rm OPT} = \sum_{n=1}^N -\log_2(c_n) \nonumber
\end{equation}
under the constraint
\begin{equation}\nonumber
	\sum_{n=1}^N \int |P_n(\omega)|^2 \mathrm{d}\omega = 2\pi N
\end{equation}
where $c_n$ is given in (\ref{cc}) and $P_n(\omega)$ is the precoder  for the channel $\Sigma_{n} (\omega)$.
By solving the Euler-Lagrange equation, the optimal precoders have spectra of the form~\eqref{eq:Pw_opt}.
\subsection{Optimal Shaping Pulse for Bandlimited AWGN Channels}\label{sec:continuous}

We now consider a linearly-modulated transmission over a continuous-time, time-invariant, bandlimited AWGN channel,
under the assumptions of ideal synchronization, and we show how to design the optimal shaping pulse for this scenario.
The received signal can be expressed as
\begin{equation}
	r(t)= \sum_k u_k \tilde{p}(t-kT) + w(t) \,, \label{eq:cont_time} \nonumber
\end{equation}
where $\tilde{p}(t)$ is the received pulse, taking into account the transmitted pulse and the channel impulse response, symbols $u_k$ are independent, zero-mean, and properly normalized such that
$\mathrm{E}\{|u_k|^2\}=1$, $T$ is the symbol time, and $w(t)$ is a zero-mean, circularly symmetric, white Gaussian noise process with
two-sided power spectral density $N_0/2$.

As before, the channel is assumed perfectly known at the receiver and time-invariant. The shaping pulse, assumed to be of unit energy, has a spectrum with support over a bandwidth $W$ and the channel frequency response is assumed flat over $W$, although the generalization to the case of a frequency-selective channel is straightforward.

A set of sufficient statistics for detection is given by the samples
at the output of a whitened matched filter (WMF)~\cite{Fo72b}, whose output has the expression (\ref{DeepEuroCrisis}) where the sequence $\{v_k\}$
has power spectral density
\begin{equation}
	|V(\omega)|^2  =  \frac{1}{T} \sum_k  \left|\tilde{P}\left(\frac{\omega}{2\pi T} -\frac{k}{T} \right) \right|^2 \nonumber 
\end{equation}
$\tilde{P}(f)$ being the Fourier transform of $\tilde{p}(t)$.
Clearly, this discrete-time model will depend on the adopted shaping pulse, its bandwidth, and the employed symbol time.

The corresponding channel shortening detector is shown in Fig.~\ref{fig:bd_awgn}b. Since the WMF can be implemented as a cascade of a continuous-time matched filter followed by a discrete-time whitening filter, this latter filter can be ``combined'' with the channel shortening filter obtaining a single discrete-time filter with frequency response~\cite{RuPr12} 
\begin{equation}\nonumber
	H^r(\omega) = \frac{G^r(\omega)+1}{|V(\omega)|^2+N_0} \, .
\end{equation}
The power spectral density of $\{v_k\}$ is
\begin{equation}\nonumber
	|V(\omega)|^2=|P(\omega)|^2|H(\omega)|^2
\end{equation}
where
\begin{equation}\nonumber
	H(\omega)= \begin{cases}
			1 & |\omega| \leq 2 WT\pi \nonumber \\
			0 & \mathrm{otherwise} \nonumber 
	           \end{cases} \,,\,\, \omega \in [-\pi,\pi] \,.
\end{equation}

Thus the optimization problem is still given by (\ref{ClownPrimeMinisters}) where the optimal shaping pulse is such that
\begin{equation}\nonumber
	|\tilde{P}(f)|^2= T |P(2\pi Tf)|^2
\end{equation}
with $|P(\omega)|^2$ given in \eqref{eq:Pw_opt}. 

Clearly, when $2 WT \geq 1$, the optimal solution is trivial and $|P(\omega)|^2$
is flat. Thus, for $2 WT=1$ the $\tilde{p}(t)$ is a $\mathrm{sinc}$ function, whereas for
$2 WT > 1$ the $\tilde{p}(t)$ can be a pulse whose spectrum has vestigial symmetry (e.g., pulses
with a root raised cosine (RRC) spectrum).
For $2WT < 1$, the  symbol time is such that the Nyquist condition for the absence of ISI cannot be satisfied.
Thus, we are working in the domain of the {\it faster-than-Nyquist} (FTN) paradigm~\cite{Ma75c,LiGe03,RuAn05} or its extension represented by time packing~\cite{BaFeCo09b,MoCoAl12}.
Note that, as said before, the discrete-time channel model, will depend on the values of $W$ and $T$. When changing the values of $W$ and/or $T$, the corresponding optimal pulse will change and so the maximum value of the AIR for the given allowed complexity. In general, when reducing the value of $WT$, the maximum AIR value will decrease. However, the spectral efficiency, defined as the ratio between the AIR and the product $WT$ could, in principle, increase~\cite{Ma75c,LiGe03,RuAn09,RuAn05,MoCoAl12,BaFeCo09b}.
This is the rationale behind FTN/time packing that allows to improve the spectral efficiency by accepting interference. The optimal value of $T$ is, in that case, properly optimized to maximize the spectral efficiency. This optimization can be now performed by also using, for each value of $T$, the corresponding optimal shaping pulse. In other words, we can find the optimal pulse for a constrained complexity detector
when FTN/time packing is adopted.

We point out that, for this scenario, the numerical computation of the optimal shaping pulse
in the time-domain can require the adoption of some windowing technique or the use of Parks-McClellan algorithm \cite{OpSc89} to obtain a practical pulse since $H(\omega)$ can have a spectrum with an ideal frequency cut.

\section{Numerical Optimization and Examples} \label{numres}
Theorem 1 provides a general form of the optimal transmit filter for
channel shortening detection of ISI channels. What remains to be
optimized is the $L+1$ complex-valued constants $\{A_{\ell}\}$. A closed
form optimization seems out of reach since the constraint
in (\ref{ClownPrimeMinisters})
has no simple analytical form in $\{A_{\ell}\}$. 

We have applied a straightforward numerical optimization of the
variables $\{A_{\ell}\}$ under the constraints
in (\ref{ClownPrimeMinisters}).
With a standard workstation and any randomly generated channel
impulse response, the optimization is stable, converges to the
same solution no matter the starting position as long as the signal-to-noise-ratio (SNR) is not very high or very low, and is altogether a matter of fractions of a second.

\begin{figure}
 \includegraphics[width=1.0\columnwidth]{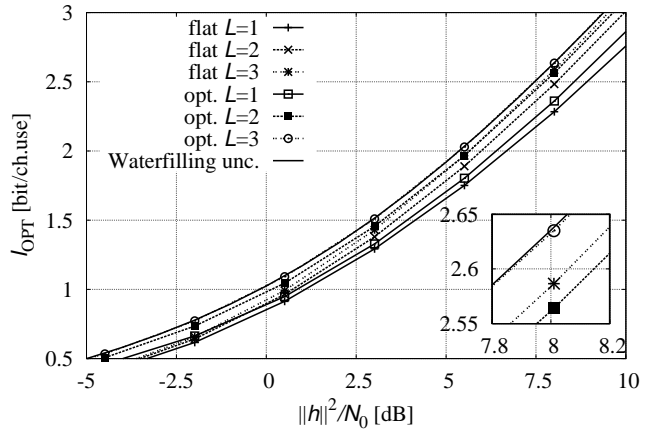}
 \vspace{-10mm}
 \caption{AIRs for Gaussian inputs when different values of the memory $L$ are considered at receiver.}\label{fig:AIR_Gauss_epr4}
 \vspace{-5mm}
\end{figure}

We now describe some illuminating examples. In all cases, the transmit power is the same both in the absence and presence of the optimal transmit filter. We first consider the complex channel $\boldsymbol{h}=[0.5,0.5,-0.5,-0.5j]$
with memory $L_H=3$.\footnote{Other examples can be found in~\cite{MoRuCo13}.}
Fig.~\ref{fig:AIR_Gauss_epr4} shows the AIR $I_{\mathrm{OPT}}$ for Gaussian
inputs when the transmit filter is optimized for different values of the memory $L$
considered by the receiver. 
For comparison, the figure also gives $I_{\mathrm{OPT}}$ for a flat transmit power spectrum  (i.e., no transmit filter at all) and
the channel capacity (i.e., when using the spectrum obtained by means of the waterfilling algorithm and assuming a receiver with unconstrained complexity). It can be seen that
using an optimized transmit filter for each $L$, 
significant gains are achieved w.r.t. the flat power spectrum
at all SNRs. The flat spectrum
reaches its maximum information rate when $L=L_H$ but suffers a loss 
to the channel capacity.
On the other hand, we can see that the optimized transmit filter 
when $L=L_H$ achieves an achievable rate which is close to the channel capacity.
However, there is not an exact match. This loss is due to the fact that 
$L_H$ must be lower than the combined channel-precoder memory $K$ as stated by Theorem~\ref{th:memory}.

\begin{figure}
  \includegraphics[width=1.0\columnwidth]{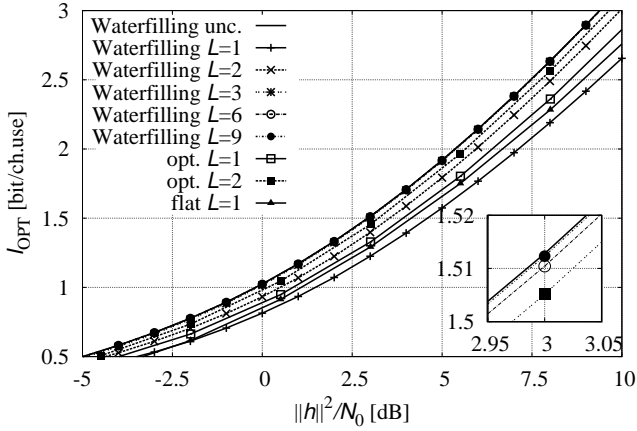}
  \vspace{-10mm}
  \caption{AIRs for Gaussian inputs with the waterfilling-solution power spectrum, when different values of the memory $L$ are considered at receiver.}\label{fig:theorem2}
\end{figure}

This behavior is clearly illustrated by Fig.~\ref{fig:theorem2}, which plots the information rate when the transmit filter is found through the waterfilling algorithm and the receiver complexity is constrained with values of the memory $L$. It can be seen that when the memory $L$ is increased more and more, even above $L_H$, the information rate becomes closer and closer to the channel capacity.
Moreover, it is important to notice 
that if, na\"{\i}vely, a transmit filter found through the waterfilling algorithm is used when the receiver complexity is constrained, 
a loss w.r.t. the optimized case occurs and it may even be better to not have any transmit filter at all for high SNR values. 

\begin{figure}
  \includegraphics[width=1.0\columnwidth]{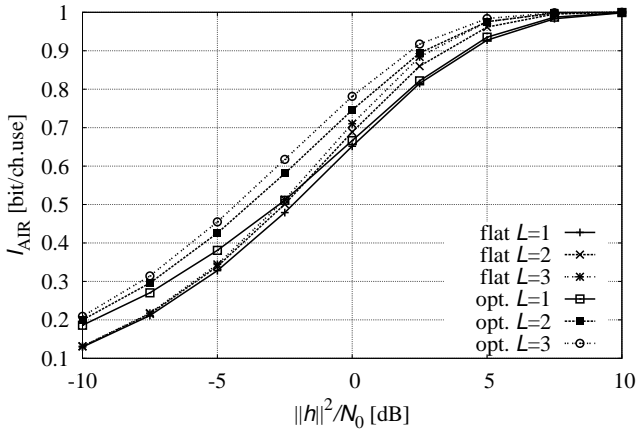}
  \vspace{-10mm}
  \caption{AIRs for BPSK modulation when different values of the memory $L$ are considered at receiver.}\label{fig:AIR_bpsk_epr4}
  \vspace{-5mm}
\end{figure}

\begin{figure}
  \includegraphics[width=1.0\columnwidth]{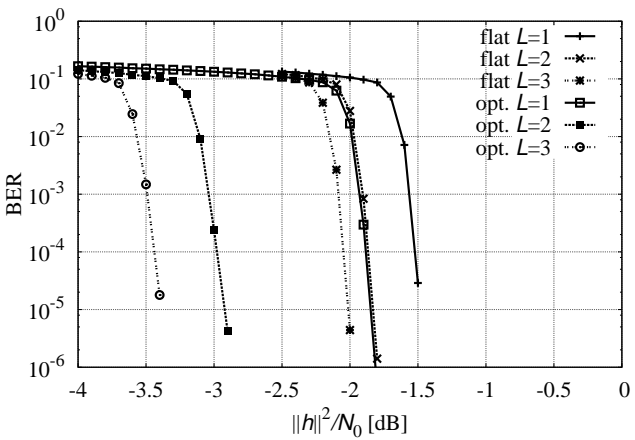}
  \vspace{-10mm}
  \caption{Bit error rate for BPSK modulation for different values of the memory $L$ considered at receiver.}\label{fig:BER_bpsk_epr4}
\end{figure}

Although the results of this paper were so far presented only for Gaussian symbols, 
we now show that when the optimized transmit filter and detector for Gaussian inputs 
are used for low-cardinality discrete alphabets, the ensuing $I_{AIR}$
is still excellent.\footnote{We remind the reader that
  $I_{\mathrm{OPT}}$ refers to an optimized detector while $I_{AIR}$ refers to the achievable rate for a non optimized detector. Since
  the filters have been optimized for Gaussian inputs, but we are
  using here low-cardinality constellations, the filters could be
  further optimized and for these reason we use the notation $I_{AIR}$.}
Fig.~\ref{fig:AIR_bpsk_epr4} shows the AIR for a binary phase shift keying (BPSK)
modulation.
It can be noticed that the behavior among the curves for BPSK  reflects the behavior for Gaussian symbols.
The AIR can be approached in practice with proper modulation and coding  formats. Fig.~\ref{fig:BER_bpsk_epr4} shows
the bit error rate (BER) of a BPSK-based system using the
DVB-S2 low-density parity-check code with rate 1/2. 
In all cases, 10 internal iterations  within the LDPC decoder and 10
global iterations were carried out. It can be noticed 
that the performance are in accordance with the AIR results.
All simulations that we have presented were also carried out
for other channels (e.g., EPR4, Proakis B and C). However, we have not presented any result for these channels since our findings for those channels are in principle identical to those for the channel presented in the paper.

\subsection{MIMO-ISI Channels with perfect CSI}
\begin{figure}
  \includegraphics[width=1.0\columnwidth]{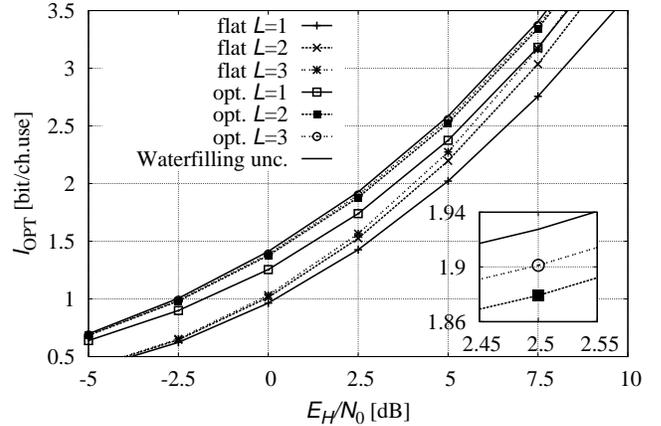}
  \vspace{-10mm}
  \caption{AIRs for Gaussian inputs over a \mbox{MIMO-ISI} channel with $N=2$ and $L_H=3$, when different values of the memory $L$ are considered at receiver.}\label{fig:AIR_MIMO_ISI}
  \vspace{-5mm}
\end{figure}

We now considered a $2\times2$ MIMO-ISI channel, with \mbox{$L_H=3$}.
Fig.~\ref{fig:AIR_MIMO_ISI} shows the AIR $I_{\mathrm{OPT}}$ for Gaussian
inputs as a function of $E_H/N_0$, being \mbox{$E_H=\sum_{\ell} \mathrm{tr}(\boldsymbol{H}_\ell \boldsymbol{H}_\ell^\dagger)$}.
The transmit filters are optimized for the equivalent channels $\Sigma_1(\omega)$ and $\Sigma_2(\omega)$
for different values of the memory $L$ considered by the receiver. 
For comparison, the figure also gives $I_{\mathrm{OPT}}$ for flat transmit power spectra (i.e., $\mathrm{E}\{\boldsymbol{a}_k \boldsymbol{a}^\dagger_{k+m} \}=\boldsymbol{I}\delta_m$, where
$\boldsymbol{I}$ is the identity matrix and $\delta_m$ is the Kronecker delta) and
the channel capacity (i.e., when using the spectra obtained by means of the waterfilling algorithm and assuming a receiver with unconstrained complexity). It can be seen that
conclusions for scalar ISI channels also hold for MIMO-ISI. However, we found that, for MIMO-ISI channel, the objective function seems to have some local maxima, and thus the optimization
can depend on the starting position. This problem can be easily solved by running the optimization more times 
(three times were always enough in all our tests) and keeping the maximum value. 

\subsection{Bandlimited AWGN channels}
\begin{figure}
	\includegraphics[width=1.0\columnwidth]{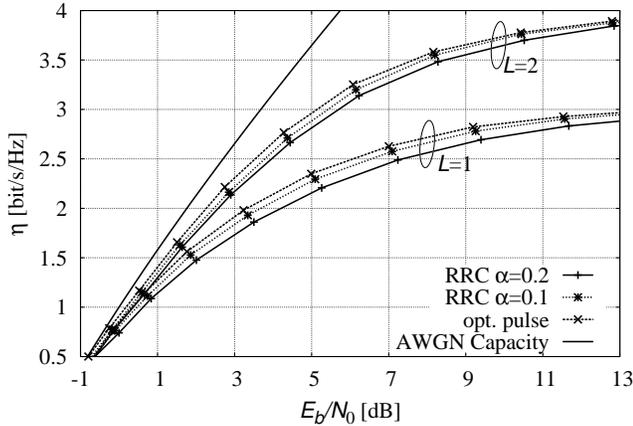}
	\vspace{-10mm}
	\caption{ASE for a BPSK modulation by using the optimized pulse for two values of the memory $L$ considered at receiver.}\label{eq:se_tpack}
\end{figure}

We computed the optimal shaping pulse on a bandlimited AWGN channel with \mbox{$2WT=0.48$}. Hence, we are in the realm of FTN/time packing and the considered ISI is only due to the adoption of such a technique.
Fig.~\ref{eq:se_tpack} shows the achievable spectral efficiency (ASE) 
\mbox{$\eta=I_{\mathrm{AIR}}/WT$} for a BPSK modulation on the \mbox{continuous-time} AWGN channel
as a function of the ratio $E_b/N_0$, $E_b$ being the received signal energy per information bit. 
Two values of the memory, namely $L=1$ and $L=2$  are considered at the detector.
For comparison, the figure also gives the ASE for pulses with RRC spectrum
and roll-off $\alpha=0.1$ or $\alpha=0.2$, 
and the unconstrained capacity for the AWGN channel.
It can be seen that the optimized pulse outperforms the other pulses.

\section{Conclusion} \label{conclusion}

We have studied ISI channels with channel shortening detection. 
The channel shortening detector that we used is optimized from a mutual information perspective and allows for the highest possible data rate. We then optimized the transmit filter for a given receiver complexity and ISI channel. 
This is an optimization problem of infinite dimensionality, but we managed to reduce it through functional analysis into an optimization problem of a dimension that equals the memory of the receiver plus one. A standard numerical optimization procedure then follows. 
Since the memory $L$ of the receiver is in practice typically set to a small value, such as $L=1$, the numerical optimization can be easily carried out.

As a side result, we also show that the classical waterfilling algorithm for ISI channels can never result in a shorter channel response at the receiver than the length of the channel response itself.
From our numerical experiments, we have found that it is crucial to take the receiver complexity into account when designing the transmit filter, since if the transmit filter found through the waterfilling algorithm is used, then a loss can occur compared with a flat transmit filter.

We have finally shown that the same framework can be used to derive the optimal shaping pulse on a bandlimited AWGN channel.

\section*{Appendix A: Proof of Theorem 1}

We first note that $P(\omega)$ only enters the
optimization through its square magnitude, and we therefore make the
variable substitution $S_p(\omega)=|P(\omega)|^2$ and optimize over
$S_p(\omega)$ instead. 

The proof will consist of three steps
\begin{itemize}
\item A formula for stationary points.
\item The observation that some of these do not have strictly positive spectrum.
\item Fixing the problem identified in the previous bullet.
\end{itemize}
Let us now start with the first bullet.

From Cramer's rule~\cite{HoJo85}, we get that 
$${\bf B}^{-1} = \frac{1}{\det({\bf B})}[C_{ij}],$$
where $C_{ij}$ is the cofactor of entry $(i,j)$ in ${\bf B}$.
This implies that we can express ${\bf b}{\bf B}^{-1}{\bf b}^{\dagger}$ as
\begin{equation} \nonumber
\frac{\sum_{m=1}^M \alpha_m
  b_0^{\phi_{m,0}}b_1^{\phi_{m,1}}(b_1^*)^{\phi_{m,2}} \cdots b_L^{\phi_{m,2L-1}} (b_L^*)^{\phi_{m,2L}} }{\sum_{n=1}^N \beta_n
  b_0^{\psi_{n,0}}b_1^{\psi_{n,1}}(b_1^*)^{\phi_{m,2}}\cdots b_{L-1}^{\psi_{n,2L-3}} (b^*_{L-1})^{\psi_{n,2L-2}}  },
\end{equation}
where $M$ and $N$ are finite constants that depend on $L$, $\alpha_m,\beta_m\in\{\pm 1\}$, and both
$\phi_{m,\ell}$ and $\psi_{n,\ell}$ are non-negative integers which
satisfy
$$\sum_{\ell=0}^{2L} \phi_{m,\ell}=L+1\quad \textrm{and}\quad \sum_{\ell=0}^{2L-2} \psi_{n,\ell}=L\,.$$

We next introduce the variable substitution 
$$y(\omega)=\frac{N_0}{|H(\omega)|^2S_p(\omega)+N_0},\,\, S_p(\omega) =\frac{N_0}{|H(\omega)|^2}\left[\frac{1}{y(\omega)}-1\right].$$
The constraint $\int S_p(\omega)\mathrm{d}\omega=2\pi$ translates into
$$e[y(\omega)]=\int_{-\pi}^{\pi} \frac{1}{y(\omega)|H(\omega)|^2}\mathrm{d}\omega =
\int_{-\pi}^{\pi}\frac{1}{|H(\omega)|^2}\mathrm{d}\omega +\frac{2\pi}{N_0}.$$
Furthermore, we have
$$b_k = \frac{1}{2\pi}\int_{-\pi}^{\pi} y(\omega) e^{jk\omega}\mathrm{d}\omega.$$
The constrained Euler-Lagrange equation~\cite{Ch10} becomes
$$\frac{\delta c}{\delta y} = \lambda \frac{\delta e}{\delta y}=-\frac{\lambda}{|H(\omega)|^2y^2(\omega)}.$$
The functional derivative $\delta b_k^s/\delta y$ equals
\begin{eqnarray}
 \frac{\delta b_k^s}{\delta y}& =& \frac{\delta \left[\int_{-\pi}^{\pi} y(\omega)e^{jk\omega}\mathrm{d}\omega\right]^s}{\delta y}\nonumber\\
				& =& s\left[\int_{-\pi}^{\pi} y(\omega)e^{jk\omega}\mathrm{d}\omega\right]^{s-1} e^{jk\omega}\nonumber \\
				& =& sb_k^{s-1}e^{jk\omega}. \nonumber
\end{eqnarray}
We now note that $b_k$, raised to any power, is a \emph{constant} that
    depends explicitly on $y$. Therefore,
    by an application on the quotient rule for the derivative and the chain rule to (\ref{cc}),
  we obtain an expression of the form
$$\frac{\delta c}{\delta y} = 1 -\frac{\sum_{\ell=-L}^L
  A_{\ell}[y]e^{j\ell\omega}}{C[y]},$$
where the constants $A_{\ell}[y]$ and $C[y]$ explicitly depend  on $y$, e.g., $$C[y]=\left[\sum_{n=1}^N \beta_n
  b_0^{\psi_{n,0}}b_1^{\psi_{n,1}}\cdots b_{L-1}^{\psi_{n,2L-3}} (b^*_{L-1})^{\psi_{n,2L-2}}\right]^2\,.$$
By manipulation of the Euler-Lagrange equation and by introducing a new set of
  constants $\{B_{\ell}[y]\}$, we obtain
$$y(\omega) =\frac{1}{\sqrt{|H(\omega)|^2[\sum_{\ell=-L}^L
    B_{\ell}[y]e^{j\ell \omega}]}}.$$
This translates into a general form of the optimal $S_p(\omega)$ which
    reads
\be \label{DegradedSerieA} S_p^{\mathrm{opt}}(\omega) =
\frac{N_0}{\sqrt{|H(\omega)|^2}}\sqrt{\sum_{\ell=-L}^L A_{\ell}e^{j\ell \omega}}-\frac{N_0}{|H(\omega)|^2}\ee
where the $A_{\ell}$ must have Hermitian symmetry.

We have now found a general form for any stationary
point. Unfortunately, for a given $H(\omega)$, this stationary point
may lie outside of the domain of the optimization. The
optimal spectrum $S_p(\omega)$ must therefore lie on the boundary of the
optimization domain, which in this case implies that \mbox{$S_p(\omega)=0$} for
$\omega\in\mathcal{I}_0\subset [-\pi,\pi].$ Let us define $\mathcal{I}_+$ as the subset $[-\pi,\pi]$ where $S_p(\omega)>0$ except for the endpoints of $\mathcal{I}_+$ where $S_p(\omega)=0$ due to the assumption of a continuous spectrum. Note that $\mathcal{I}_+$  may be the union of several disjoint sub-intervals of $[-\pi,\pi]$. We can now rewrite the constraint and the expressions of $b_k$ as

$$e[y(\omega)]=
\int_{\mathcal{I}_+}\frac{1}{|H(\omega)|^2}\mathrm{d}\omega +\frac{2\pi}{N_0}$$
and
$$b_k = \frac{1}{2\pi}\int_{\mathcal{I}_+} y(\omega) e^{jk\omega}\mathrm{d}\omega.$$
From the first part of the proof, i.e., identifying a necessary condition for stationary points, we have that 
(\ref{DegradedSerieA}) must hold within the interval $\mathcal{I}_+$, and the constants $\{A_{\ell}\}$ must be such that $ S_p^{\mathrm{opt}}(\omega)=0$ at the end-points of each sub-interval within $\mathcal{I}_+$. Hence, no matter what $\mathcal{I}_+$ is,  we can express the optimal $S_p^{\mathrm{opt}}(\omega)$ as in (\ref{eq:Pw_opt}).


\section*{Appendix B: Proof of Theorem 2}

The waterfilling algorithm provides a
transmit filter that satisfies \cite{Hi88}
\be \label{FotballScandals}|P(\omega)|^2 = \max\left(0,\theta-\frac{N_0}{|H(\omega)|^2}\right),\ee
for some power constant $\theta$. In view of Theorem 1,
$|P(\omega)|^2$ in (\ref{FotballScandals}) must also satisfy (\ref{eq:Pw_opt}).
Equating (\ref{FotballScandals}) and (\ref{eq:Pw_opt}) yields
\be \label{CostaConcordia} \theta-\frac{N_0}{|H(\omega)|^2}=\frac{N_0}{\sqrt{|H(\omega)|^2}}\sqrt{\sum_{\ell=-K}^K A_{\ell} e^{j\ell \omega}}-\frac{N_0}{|H(\omega)|^2}.\ee
From (\ref{CostaConcordia}), it can be seen that we must have 
$$\sum_{\ell=-K}^K A_{\ell} e^{j\ell \omega} = \gamma |H(\omega)|^2,$$
for some constant $\gamma$. However, 
$$ |H(\omega)|^2 = \left|\sum_{\ell=0}^{L_{\mathrm{H}}} h_\ell e^{-j\ell \omega} \right|^2= \sum_{\ell=-L_{\mathrm{H}}}^{L_{\mathrm{H}}} g_{\ell}
e^{-j\ell \omega}, $$
where 
$$g_\ell=\sum_k {h_k h^*_{k-\ell}}.$$
Clearly,  to satisfy
$$\sum_{\ell=-K}^K A_{\ell} e^{j\ell \omega} =\gamma\left[\sum_{\ell=-L_{\mathrm{H}}}^{L_{\mathrm{H}}} g_{\ell}e^{-j\ell \omega}\right],$$
$K$ must at least equal  $L_{\mathrm{H}}$.

\begin{biography}[{\includegraphics[width=1in,height=1.25in,clip,keepaspectratio]{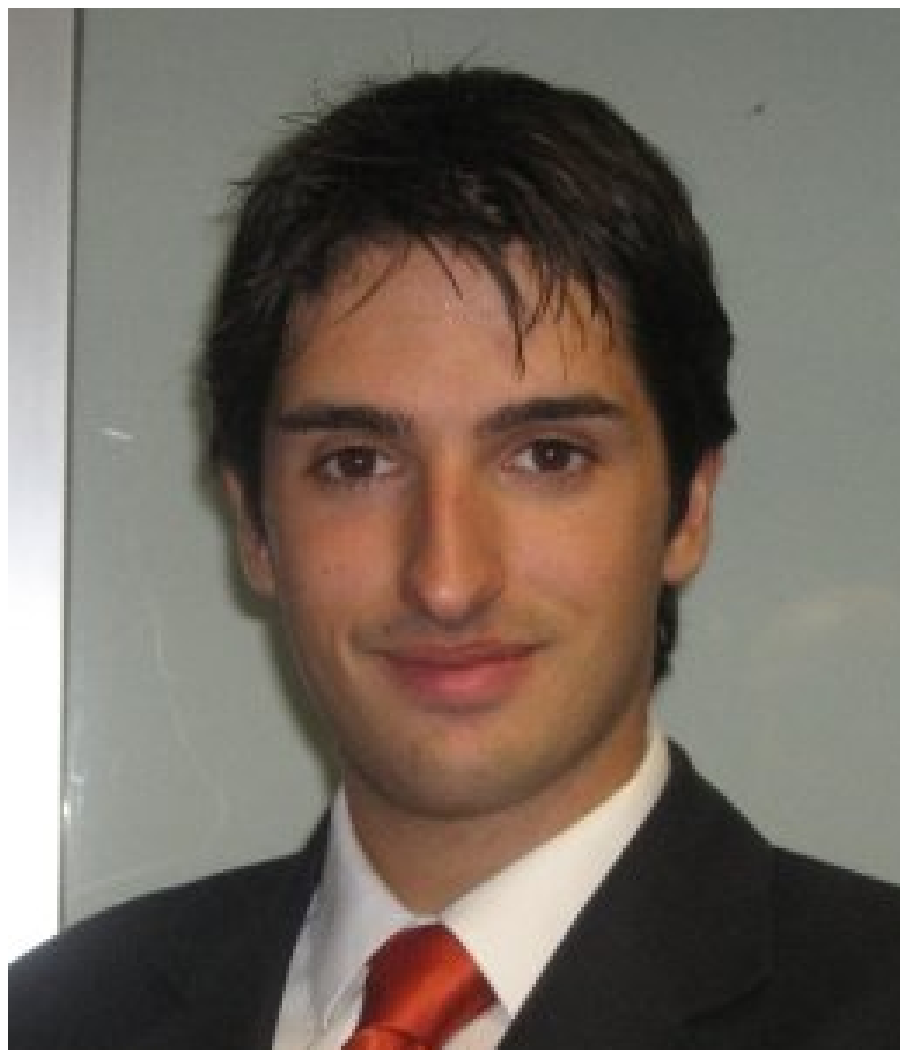}}]
{Andrea Modenini} (S'12) was born in Parma, Italy, in 1986. He received the Dr. Eng. degree in telecommunications engineering (cum laude) in december 2010 from the University of Parma, Italy, where he is currently Ph.D. Student at the Dipartimento di Ingegneria dell'Informazione (DII). His main research interests include information theory and digital transmission theory, with particular emphasis on the optimization of detection algorithm from an information theoretic point of view.
He participates in several research projects funded by the European Space Agency (ESA-ESTEC) and important telecommunications companies. In the spring 2012 he was a visiting PhD student at the University of Lund, Sweden, for reasearch on channel shortening detection for spectrally efficient modulations.
\end{biography}
\vspace{-3mm}
\begin{biography}[{\includegraphics[width=1in,height=1.25in,clip,keepaspectratio]{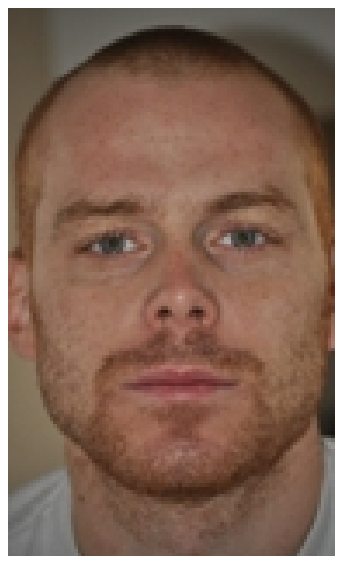}}]
{Fredrik Rusek} was born in Lund, Sweden on April 11, 1978. He received the
Master of Science degree in electrical engineering in December 2002 and the Ph.D.
degree in digital communication theory in September 2007, both from Lund
Institute of Technology. In October 2007 he joined the the department of electrical
and information technology at Lund Institute and since 2012, he holds an associate professorship at the same department. Since September 2012, he is also part time employed as algorithm specialist at Huawei Technologies, Lund, Sweden.
His research
interests include modulation theory, equalization,
wireless communications, and applied information theory.
\end{biography}
\vspace{-3mm} 
\begin{biography}[{\includegraphics[width=1in,height=1.25in,clip]{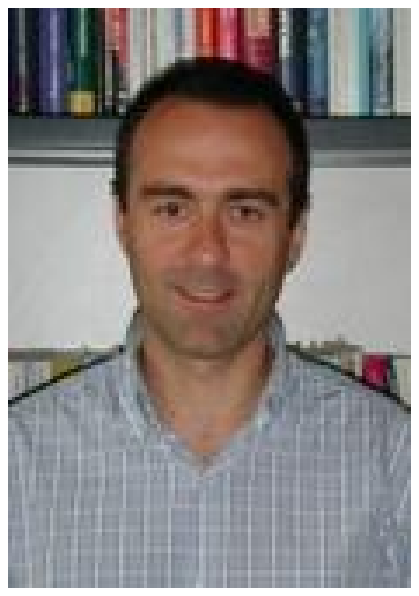}}]
{Giulio Colavolpe} (S'96-M'00-SM'11) was born in Cosenza, Italy, in 1969. He received the Dr. Ing. degree in Telecommunications Engineering (cum laude) from the University of Pisa, in 1994 and the Ph.D. degree in Information Technologies from the University of Parma, Italy, in 1998. Since 1997, he has been at the University of Parma, Italy, where he is now an Associate Professor of Telecommunications. In 2000, he was Visiting Scientist at the Institut Eur\'ecom, Valbonne, France. His research interests include the design of digital communication systems, adaptive signal
processing (with particular emphasis on iterative detection techniques for channels with memory), and information theory.

He received the best paper award at the 13th International Conference on Software, Telecommunications and Computer Networks (SoftCOM'05), Split, Croatia, September 2005, the best paper award for Optical Networks and Systems at the IEEE International Conference on Communcations (ICC 2008), Beijing, China, May 2008, and the best paper award at the 5th Advanced Satellite Mobile Systems Conference and 11th International
Workshop on Signal Processing for Space Communications (ASMS\&SPSC 2010), Cagliari, Italy.  He is currently serving as an Editor for \textit{IEEE Transactions on Communications} and \textit{IEEE Wireless Communications Letters}. He also served as an Editor for \textit{IEEE Transactions on Wireless Communications} and as an  Executive Editor for \textit{Transactions on Emerging Telecommunications Technologies (ETT)}.
\end{biography}
 
\end{document}